\documentclass[12pt]{article}

\usepackage{epsfig}

\def\hybrid{\topmargin -20pt  \oddsidemargin 0pt
      \headheight 0pt   \headsep 0pt
      \textwidth 6.25in 
      \textheight 9.5in 
      \marginparwidth .875in
      \parskip 5pt plus 1pt   \jot = 1.5ex}

\hybrid

\usepackage{amsfonts}
\usepackage{amssymb}

\begin{document}

\def\x{\times}
\def\pa{\partial}
\def\ra{\rightarrow}
\def\lra{\leftrightarrow}
\def\beq{\begin{equation}}
\def\eeq{\end{equation}}
\def\beqa{\begin{eqnarray}}
\def\eeqa{\end{eqnarray}}

\sloppy
\newcommand{\Tr}{{\rm Tr}}
\newcommand{\tr}{{\rm tr}}
\newcommand{\be}{\begin{equation}}
\newcommand{\eq}{\end{equation}}
\newcommand{\non}{\\ \nonumber}
\newcommand{\dd}{ {\rm d} }
\renewcommand{\arraystretch}{1.5}

\renewcommand{\thesection}{\arabic{section}}
\renewcommand{\theequation}{\thesection.\arabic{equation}}

\parindent0em

\begin{titlepage}
\begin{center}
\hfill HU-EP-01/23\\
\hfill {\tt hep-th/0106155}\\

\vspace{2cm}

{\LARGE {\bf Fluxes in Heterotic and \\[.3cm]

Type II String Compactifications}}

\vspace{1cm}

{\bf Gottfried Curio, Albrecht Klemm, Boris K\"ors and 
Dieter L\"ust}\footnote{\mbox{email: \tt 
curio,aklemm,koers,luest@physik.hu-berlin.de}}\\
\vskip 1cm

{\em Humboldt-Universit\"at zu Berlin,
Institut f\"ur Physik, \\ 
D-10115 Berlin, Germany}
\vskip .1in

\end{center}

\vspace{1.5cm}

\begin{center} {\bf ABSTRACT } \end{center}
\begin{quotation}\noindent
In this paper we consider heterotic compactifications on $K3\times
\mathbb{T}^2$ as well as type II compactifications on $K3$-fibred
Calabi-Yau spaces with certain fluxes for the gauge and RR field strengths
F and H turned on. By providing an identification of corresponding fluxes 
we show that the well-known ${\cal N}=2$ heterotic/type II string-string
duality still holds for a subset of all possible fluxes, 
namely those which arise from six-dimensional gauge fields 
with internal magnetic flux on the common two-sphere ${\mathbb P}_b^1$, 
which is the base space of the type II $K3$-fibration. 
On the other hand, F- and H-fluxes 
without ${\mathbb P}_b^1$-support, such as heterotic F-fluxes
on the torus ${\mathbb T}^2$ or type II H-fluxes on cycles of the $K3$-fibre
cannot be matched in any simple way, which is a challenge for 
heterotic/type II string-string duality.
Our analysis is based on the comparison of terms in the effective 
low-energy heterotic and type II actions which are induced by the fluxes, 
such as the Green-Schwarz couplings related to flux-induced $U(1)$ anomalies, 
the effective superpotential and the Fayet-Iliopoulos scalar potential.

\end{quotation}

\end{titlepage}
\vfill
\eject

\newpage

\newpage
\section{Introduction}

String compactifications with background fluxes constitute an interesting
class of string vacua.
In type IIA/B compactifications on Calabi-Yau three-folds background fluxes
are provided by vacuum expectation values of internal NS and R 
H-fields \cite{PolStro,Mich,Gukov:2000ya,Gukov:2000gr,VT,Mayr,CKLTh}.
In the local case a dual description of closed type II strings on
Calabi-Yau spaces with H-fluxes was proposed in terms of open topological 
strings, corresponding to a duality of superstrings on non-compact Calabi-Yau spaces 
with H-fluxes and certain large N gauge systems with ${\cal N}=1$ supersymmetry 
in four dimensions
\cite{Vafa:2000wi}.
In ${\cal N}=2$ heterotic string compactifications on $K3\x \mathbb{T}^2$ vacua
with background fluxes can be obtained by turning on
internal magnetic fields (magnetic F-fluxes) on certain
internal
two-dimensional subspaces 
\cite{Tseytlin,Anton}.\footnote{Type I
compactifications with internal
magnetic fluxes on tori have also been constructed 
\cite{Bachasal,Blumenhagen:2000wh,Angelantonj:2000hi,Blumenhagen:2001vk,Angelantonj:2000rw,Aldazabal:2000dg,Blumenhagen:2001ea,Ibanez:2001nd}.}

In general, the heterotic F-fluxes or, respectively, the type II
H-fluxes cause several interesting effects 
after compactification to four dimensions: they can introduce
warped space-times \cite{Verlinde:2000fy,Curio:2001dw,Giddings:2001yu},
they lift the vacuum degeneracy
in the moduli space, they induce a spontaneous breaking of ${\cal N}=2$
supersymmetry, where only at special points in the moduli space
supersymmetry may be unbroken; they can imply
spontaneous 
breaking of $U(1)$ gauge symmetries, they may create tachyonic scalar
fields, 
which destabilize the vacuum, and they generate chiral fermion spectra. 
All these effects are in principle  of vital phenomenological interest. 

Based on the ${\cal N}=2$ heterotic/type II string duality, which was
established
in the absence of background fluxes \cite{Kachru:1995wm}, 
we will discuss in how far this
duality still holds when heterotic F-fluxes resp. type II H-fluxes are
turned on. In particular we will show how at least
part of these fluxes will be
mapped onto each other via the heterotic/type II string duality.
However other heterotic F-fluxes or type H-fluxes find no obvious
dual interpretation and are therefore in apparent conflict with the
heterotic/type II string-string duality.
Our discussion will be mainly based on the comparison of several terms in
the four-dimensional
effective action:

\begin{itemize}

\item 
In type
IIA on a Calabi-Yau
space $M$ the
internal $H_R^{(n)}$-fluxes 
generate a  moduli dependent superpotential \cite{VT}
\beqa
W = \sum_{n=0}^3\int_M H_R^{(2n)}\wedge J^{3-n}= 
e_I~ X^I(z) - m^I~
{\cal F}_I(z)\, ,
\eeqa 
where $J$ is the K\"ahler class and 
$(X^I,{\cal F}_I)$ ($I=0,\dots ,N_V$) is the symplectic vector of ${\cal N}=2$
special geometry which depends
on the 
scalar fields $z^A$ ($A=1,\dots ,N_V$) in the vector multiplets.
This superpotential generically breaks supersymmetry and the masses of
the gravitini are of the order  $M_{3/2}\simeq |W|^2$.
In addition the ground state of the corresponding potential determines
in general the values of the complex scalar fields $z^A$.
The $e_I$ and $m^I$ are the quantized values of the H-fluxes on the
0,2,4 and 6-cycles of $M$, namely
$e_0=\int H_R^{(6)}$, $e_A=\int H_R^{(4)}$, $m^A=\int H_R^{(2)}$,
$m^0=\int H_R^{(0)}$. 
We will show that those type IIA
H-fluxes, which have support on a certain two-sphere ${\mathbb P}^1$,
correspond in the heterotic string to gauge F-fluxes on the `same'
${\mathbb P}^1$. 
These are essentially $e_0$ and all $e_A$, except the one without
${\mathbb P}^1$ support.
On the other hand,  IIA H-fluxes which no
support on ${\mathbb P}^1$, namely all except one $m^A$ as well as
$m^0$, have no immediate heterotic interpretation.

\item
In the heterotic string internal F-fluxes on some two-cycle $C_i$ induce
four-dimensional Green-Schwarz couplings via the ten-dimensional
Chern-Simons term $B_2\wedge \Tr (F^4)$ as well as through the  
Chern-Simons interactions in $H\wedge *H$. They involve 
the $U(1)$ vector fields $A^I$ and 
the internal B-field  $B=a_{C_i} J_{C_i}+\ldots$, where $C_i$ will be  either
a certain  2-sphere ${\mathbb P}^1$ or a certain 2-torus ${\mathbb T}^2$. The 
four-dimensional couplings proportional to internal fluxes $e^i_I$ of
$F^I=dA^I$ on $C_i$ are of the form ($ *_4 d a_{C_i}=:d B_{C_i}$) 
\beqa
{\cal L}_{GS} = e_I^i ~B_{C_i} \wedge F^I .
\eeqa
These couplings are responsible for the longitudinal component 
of a $U(1)$ vector boson which becomes massive due to the
fluxes \cite{DSWW,DSW}. In addition the Green-Schwarz term
can cancel possible $U(1)$ triangle anomalies due to massless chiral
fermions (see below). For the case of ${\mathbb P}^1$-fluxes
the same Green-Schwarz action is
obtained from the ten-dimensional type IIA action turning on
H-fluxes with  ${\mathbb P}^1$ support. In fact, the Green-Schwarz action
gives a very direct mean to map the heterotic gauge F-fluxes to the
type H-fluxes and vice versa.

\item
Along with this Green-Schwarz coupling there will be in general a mass
term of the $U(1)$ gauge bosons,
\beqa
M_{A^I}^2\sim e_I^2\, ,
\eeqa
which signals a spontaneous gauge symmetry breaking due to the fluxes
$e_I$. In the heterotic string  this term comes from the ten-dimensional
kinetic term for the CS-improved $H$. 
However in the type II compactifications the
same term is not present in the tree-level effective action, but comes as 
a one-loop effect.

\item
The two-dimensional index theorem for the Dirac operator 
relates the net number
of chiral fermions to the fluxes $e_I$. These fermions are in general
charged
under the corresponding $U(1)$ gauge group, and hence
a $U(1)$ triangle anomaly may be implied. It will
be canceled by the Green-Schwarz term described above, together
with a second coupling of the form
\beqa
{\cal L}=c_I~ a_{C_i}\, F^I\wedge F^I\, ,
\eeqa
where the $c_I$ denote the strength of this interaction. Then the chiral anomaly is given as
the product $e^i_Ic_I$. For non-zero fluxes
$e_I$ and  non-vanishing Green-Schwarz term the chiral anomaly is nevertheless
absent, provided the coefficients $c_I$ are zero. 
In other words, the mass generation for the $U(1)$ gauge boson 
by the Green-Schwarz term is not
necessarily linked to a non-trivial $U(1)$ anomaly.
Precisely this, non-vanishing Green-Schwarz couplings without anomaly, 
will happen for the
${\mathbb P}_b^1$-fluxes. On the contrary, chiral anomalies are in general
present for heterotic ${\mathbb T}^2$-fluxes, which has also been confirmed in
a large class of type I compactifications with ${\mathbb T}^2$-fluxes
(see also the discussion about $U(1)$ anomalies in \cite{Ibanez:2001nd}).

\item
In general, a Fayet-Iliopoulos (FI) D-term scalar
potential will be generated
\cite{Lerche}, depending on the fluxes $e_I$, which contains tachyonic mass
terms for charged scalar fields.

\end{itemize}

In our paper we will consider on the heterotic side two equivalent
string compactifications to four dimensions, which proceed however via an
inequivalent 
intermediate six-dimensional compactification.
In compactification scheme A the $E_8\x E_8$ heterotic string is
compactified from
ten to six dimensions on a $\mathbb{T}^2_{f}\x \mathbb{T}^2_c$. The
six-dimensional theory is 
then  fibred over a base $\mathbb{P}^1_b$, where the first torus
$\mathbb{T}^2_{f}$ varies over the $\mathbb{P}^1_b$ to build the $K3$
and the second one
is constant. This will provide 20 perturbative Abelian gauge
multiplets $A_{\mu}$ already in six dimensions from which one can build
the magnetic F-fluxes $\int_{\mathbb{P}^1_b}dA_{\mu}$ on
$\mathbb{P}^1_b$. 

On the other hand, in the heterotic string one can
also have fluxes on the $\mathbb{T}^2_c$; then
one needs to have a gauge field in 
six dimensions from a compactification of the heterotic string on $K3$,
i.e. in compactification 
scheme B one compactifies from ten to six dimensions on $K3$ and then
from six to four dimensions 
on a $\mathbb{T}^2_c$. In order to have Abelian gauge fields in six
dimensions, the moduli of the gauge bundle on $K3$ 
must be tuned to specific values.

On the dual type IIA side we will deal with a Calabi-Yau space $M$ which
is
a K3-fibration \cite{Klemm:1995tj,Aspinwall:1996vk}
over the same base $\mathbb{P}^1_b$. 
Using the adiabatic principle we will be able to completely map
the IIA RR H-fluxes with support on
${\mathbb P}^1_b$ to the magnetic F-fluxes on the heterotic side, again
on the `same' ${\mathbb P}^1_b$. Note that
it was argued already in \cite{PolStro}  that the IIA 6-flux on $M$,
corresponding to 
$e_0$, is mapped under the six-dimensional string-string duality to a
magnetic field on the 
heterotic torus. We will find however that the type IIA H-fluxes
investigated in \cite{VT} and \cite{CKLTh} are mapped to heterotic
fluxes on $\mathbb{P}^1_b$, cf. \cite{Anton}.
However the heterotic F-fluxes on the
$\mathbb{T}^2$ are of a different nature and cannot be simply transfered
to the type IIA side.
In the other direction the type IIA H-fluxes on $K3_{IIA}$ have no simple
heterotic interpretation. These H-fluxes lead to a massive IIA supergravity
theory in six dimensions, for which the heterotic/type IIA duality
apparently does not work \cite{Haack:2001iz}.
Finally, in type IIB compactifications there are also fluxes from the
NS 3-form field $H_{NS}^{(3)}$. These H-fluxes are already difficult to
understand in type IIA.\footnote{In 
\cite{Vafa:2000wi} it was argued that the type IIB  
$H_{NS}^{(3)}$-fluxes might correspond
to NS 4-form fluxes in the dual type IIA models.} Also in the heterotic
models these fluxes have no obvious interpretation.

The paper is organized as follows. In section \ref{revhet} we review 
the ${\cal N}=2$
theory in four dimensions, which emerges from from the heterotic
compactification on $K3\times \mathbb{T}_c^2$.  
In section
\ref{sixduality} we collect those  
facts about the six-dimensional duality that are  relevant to the precise 
$ \mathbb{P}_b^1$ flux mapping, which is presented in section \ref{p1fluxmap}.
In particular  the identification of the fluxes can be done
by comparing the Chern-Simons 
terms, as discussed in \ref{gsterms}. We discuss in some detail the
four-dimensional  terms induced in the 
superpotential from the presence of non-trivial heterotic F-fluxes 
\ref{hetsupo}. The heterotic $\mathbb{T}^2_c$ fluxes
are discussed in 
section \ref{hett2fluxes}. 
In section 5 we discuss the possible emergence of tachyons due to 
fluxes. The paper ends with some conclusions.

\section{Type II/heterotic string duality without fluxes}   

\subsection{Heterotic strings on $K3\times \mathbb{T}_c^2$ and type IIA on $M$}
\label{revhet}
  
Let us recall  the spectrum of the $E_8\times E_8$ heterotic string 
compactified on $K3\times \mathbb{T}^2_c$ (for a review on ${\cal N}=2$
string
compactifications see \cite{lercherev,louis,klemm,lust}).
It consists out of the ${\cal N}=2$ gravity multiplet, containing the
graviphoton field $\gamma$, $N_V$ vector multiplets $z^A$ and 
$N_H$ hyper multiplets $q_i$. The following three $U(1)$ vector fields do
not depend 
on the specific gauge bundle of $K3$ and are therefore universal. 
The heterotic dilaton is the scalar component in a vector multiplet
$S$.  
Further the complexified K\"ahler class and the complex structure of $\mathbb{T}^2_c$
are scalar 
components in the vector multiplets $T$ and $U$. 
In addition there are in general 20 hyper multiplets, namely the
moduli of $K3$, one of these contains the volume of the
base $ \mathbb{P}^1_b$ and will be denoted ${\rm vol}({\mathbb{P}^1_b})_{het}$.
In fact, one can consider situations, i.e. gauge bundles, where on the
vector
side only the fields $\gamma$, $S$, $T$ and $U$ are present
($N_V=3$). The corresponding
classical
(i.e. $S\rightarrow\infty$) ${\cal N}=2$ prepotential ${\cal F}$
then takes the form:
\beqa
{\cal F}= -i(X^0)^2~S ~T~U\, ,\label{hetprepot}
\eeqa
where $S=-iX^1/X^0$, $T=-iX^2/X^0$ and $U=-iX^3/X^0$ and $X^0$
corresponds
to the graviphoton.
This prepotential determines the gauge couplings of the $U(1)^4$
gauge group via these general formulas:
\begin{equation}
{\cal L}^{\rm gauge}\ =\ -{\textstyle{i\over 8}}
\left( {\cal N}_{IJ}\,F_{\mu\nu}^{+I}F^{+\mu\nu J}\
-\ \bar{\cal N}_{IJ}\,F_{\mu\nu}^{-I} F^{-\mu\nu J} \right)\, ,
\end{equation}
where $F^{\pm I}_{\mu\nu}$ ($I=1,\dots ,N_V$)
denote the self-dual and anti-self-dual 
electric field-strength components and 
\begin{equation}
{\cal N}_{IJ}=
\bar{\cal F}_{IJ}+2i~ {{\rm Im}({\cal F}_{IK})
{\rm Im}({\cal F}_{JL})X^KX^L\over
({\cal F}_{KL}) X^KX^L}\, .
\label{Ndef}
\end{equation}
Hence ${\cal N}_{IJ}$ is the field-dependent tensor that comprises the 
inverse gauge couplings
$g^{-2}_{IJ}= {i\over 4}({\cal N}_{IJ}-\bar {\cal N}_{IJ})$
and the generalized $\theta$ angles
$\theta_{IJ}= 2\pi^2({\cal N}_{IJ}+\bar {\cal N}_{IJ})$. 
We can also define magnetic field strength tensors
$G^\pm_{\mu\nu I}$ as
\begin{equation}
G^+_{\mu\nu I}={\cal N}_{IJ}F^{+J}_{\mu\nu},\quad G^-_{\mu\nu I}=
\bar{\cal N}_{IJ}F^{-J}_{\mu\nu}\, .\label{magneticg}
\end{equation}
The duality group acts on the field strength vectors 
$(F^{+J}_{\mu\nu}, G^{+}_{\mu\nu I})$ as $Sp(2N_V+2,\mathbb{Z})$
transformation
\beqa
\hat F^{+I}_{\mu\nu}& =& U^I_J F^{+J}_{\mu\nu}+Z^{IJ}
G^+_{\mu\nu J}\, , \nonumber\\
\hat G^+_{\mu\nu I}& = & W_{IJ}F^{+J}_{\mu\nu}+V_I^JG^+_{\mu\nu J} \, 
\label{dualsympa}
\eeqa
and similarly on the period vector $(X^I,{\cal F}_J)$
\beqa 
\hat X^I  &=& U^I_JX^I+Z^{IJ}{\cal F}_J,\cr
\hat{\cal F}_I &=& W_{IJ}X^J+  V_I^J{\cal F}_J\ . \label{dualsympb} \eeqa
This induces the following transformations on the gauge coupling constants
\beqa
\hat{\cal N}_{IJ} & =& (V_I^K{\cal N}_{KL}+W_{IL})\lbrack (U+Z{\cal
N})^{-1}
\rbrack^L_J\, .
\label{dualsympc}
\eeqa

{}From the heterotic string theory we know that {\em all}
the {\em physical} low-energy couplings become weak in the 
large-dilaton limit, which suggests that the strongly-coupled 
$F_{\mu\nu}^{+S}$ field
strength in the dilaton ${\cal N}=2$ superfield should be replaced
with its dual (which is weakly coupled in the large-dilaton limit).
In ${\cal N}=2$ terms, this is achieved by the symplectic transformation
 $\hat X^1 =  {\cal F}_1$,  $\hat{\cal F}_1 = -  X^1$.
The gauge couplings in the new basis is \cite{dWKLL}
\beqa
\hat{{\cal N}}_{IJ} = -2i \bar S\, \eta_{IJ} +2i(S + \bar S)\,
{\eta_{IK}\,\eta_{JL}(\hat{z}{}^K \hat{\bar z}{}^L + \hat{\bar 
z}{}^K \hat{z}{}^L)
\over  \hat{z}{}^K \eta_{KL} \hat{\bar z}{}^L} \, .
\label{goods}
\eeqa
Note that now all ${\rm Im}~\hat{\cal N}_{IJ}$
are proportional to $S+\bar S$ and hence {\em all} the gauge
couplings become weak in the large-dilaton limit.
Among the duality transformations (\ref{dualsympc}) acting on 
$(\hat X^1,\hat{\cal F}_1)$ we distinguish the perturbative 
T-dualities, which do not mix $\hat X^I$ with $\hat{\cal F}_I$, 
in contrast to the non-perturbative electro-magnetic S-duality 
transformations, which do mix $\hat X^I$ with $\hat{\cal F}_I$ (see 
\cite{dWKLL,LopesCardoso:1995zq,LopesCardoso:1996nk} for more details).

On the dual type IIA side the massless spectrum is determined
by the cohomology of the three-fold $M$, namely by the two Hodge numbers
$h^{(1,1)}$ and $h^{(2,1)}$. Specifically, the number
of vector multiplets is $N_V=h^{(1,1)}$. 
Their $h^{(1,1)}$ complex scalar moduli in the NS-NS
sector correspond to the deformations of the K\"ahler form $J$ of
$M$ plus the internal
$B_{MN}$ fields; the $h^{(1,1)}$ $U(1)$ R-R vectors  originate from
the ten-dimensional 3-form gauge potential $A_{MNP}$ with two indices
in the internal space. Second, 
there are $N_H=h^{(2,1)}+1$ massless ${\cal N}=2$
hyper multiplets.
$h^{(2,1)}$ of them correspond to the complex structure
deformations of $M$, where the two additional R-R
scalar degrees of freedom, needed to fill an ${\cal N}=2$ hyper multiplet,
come from $A_{MNP}$ with all indices in the internal direction. The
additional
hyper multiplet contains together with the
NS-NS axion field $a$ the four-dimensional dilaton $e^{-2\phi_{IIA}}$
plus two  more R-R scalar fields.

As stated already we assume that $M$ is $K3_{IIA}$ fibered as well as elliptically
fibered by the elliptic fibre $f$ (which is also the elliptic fibre of
the elliptically fibered $K3_{IIA}$). In accordance with our
earlier assumption that on the heterotic side there are just the vector
multiplets $S$, $T$, $U$ and $\gamma$ present,
we consider $M$ to be one of the 
three $(3,243)$ CY's over the Hirzebruch surfaces $F_0$, $F_1$, $F_2$
related to the heterotic instanton numbers (12,12), (11,13), (10,14) \cite{MV}.
The first of these is elliptically fibred 
over $ \mathbb{P}^1_b \times \mathbb{P}^1_f$ and has the further
heterotic/heterotic
duality related to the exchange of the two $\mathbb{P}^1$'s,
the second one was studied in connection with the five-brane
transition and its geometric counterpart of blowing down a del Pezzo
surface lying in the elliptic fibration over $\mathbb{P}^1_b$,
and finally the last one is the Calabi-Yau hypersurface of 
degree 24 in the weighted projective space ${\bf P}_{1,1,2,8,12}$.
Let us denote the base and fibre $ \mathbb{P}^1$ of the base surface
$F_n$
respectively by $ \mathbb{P}^1_b$ and $\mathbb{P}^1_f$.
Having Hodge numbers
$h^{(1,1)}=3$ there are just three K\"ahler moduli, corresponding to 
the volumes of $ \mathbb{P}^1_b$, $\mathbb{P}^1_f$ and the elliptic
fibre
$\mathbb{T}^2_f$, respectively.
Via the heterotic/type IIA string duality these fields can be mapped to
the heterotic fields $S$, $T$ and $U$ as follows:
\begin{equation}
S\sim{\rm vol}({\mathbb{P}^1_b})_{IIA},\quad T\sim{\rm
vol}({\mathbb{T}^2_f})_{IIA},\quad
U\sim {\rm vol}({\mathbb{P}^1_f})_{IIA}\, .\label{hetiimap}
\end{equation}
Note that this is the correct identification for the model with
instanton number (12,12) which corresponds to the Hirzebruch base $F_0$.
For $F_1$ and $F_2$ the identification of ${\rm vol}({\mathbb P}^1_f)_{IIA}$
is slightly different, e.g. for $F_2$ one has
$U-T\sim{\rm vol}({\mathbb P}^1_f)_{IIA}$.
On the other hand, the  type II dilaton, which is inside the 
universal hyper multiplet, is mapped to the volume of the heterotic
$ \mathbb{P}^1_b$:
\begin{equation}
e^{-2\phi_{IIA}} \sim{\rm vol}({\mathbb{P}^1_b})_{het}\, .
\label{iihetmap}
\end{equation}

\subsection{The six-dimensional duality}
\label{sixduality}

Following \cite{Sen} and \cite{Duff} we compare the actions in six
dimensions
for the heterotic string on $\mathbb{T}^4$ and the type IIA string on
$K3$.
Consider first the heterotic string on
$\mathbb{T}^4=\mathbb{T}^2_{f}\times \mathbb{T}^2_c$.
Then the vector fields $\gamma$, $S$, $T$ and $U$ are already present
in six dimensions. They correspond to the left- and right-moving
gauge group $U(1)_L^2\times U(1)_R^2$ which is associated to
the torus $\mathbb{T}^2_c$. In addition there are
$4$ $U(1)$ vectors from $\mathbb{T}^2_f$ and $16$ from the gauge group
in ten dimensions.
Including the Chern-Simons terms in the field strength of the heterotic $B$
field,
\beqa
H_{\mu\nu\rho}=(\pa_{\mu}B_{\nu\rho}+2A_{\mu}^IL_{IJ}F^J_{\nu\rho})
+{\rm cyclic}\, ,
\eeqa
one finds for the heterotic string action
\beqa\label{sixdhetaction}
S_{het}&\sim& \int d^6x \sqrt{-g_{het}}e^{-\phi_{het}} 
\Bigl( R_{het} + (\pa_{\mu}\phi_{het})^2
-\frac{1}{2\cdot 3!}H_{het}^2 \\
& & \hspace{4cm} +\, \frac{1}{8}{\rm Tr}(\pa_{\mu}M_{het}L\pa_{\mu}M_{het}L)
-F^I_{het}(LM_{het}L)_{IJ}F^J_{het}\Bigr) \nonumber
\eeqa
where $F^I_{het}$ 
($I=1,\dots , 24$) are the $24$ abelian gauge field strengths  and 
$M_{het}$ is the (symmetric $24\times 24$ matrix valued) scalar field
representing an
element of the $O(4,20)$ coset with $M_{het}LM_{het}^T=L$ for the intersection
form
\beqa
L={\left(\begin{array}{cc}I_4&\\&-I_{20}\end{array}\right)} .
\eeqa
Note that the $T,U$ fields are not in a 
diagonal basis but in one with the $H$ intersection form
consisting of a hyperbolic plane.
Upon further compactification  on  $\mathbb{P}^1_b$
the $U(1)$ gauge  fields $A^\gamma_\mu$, $A^S_\mu$, $A^T_\mu$ and
$A^U_\mu$
will always be present in four dimensions. Then the matrix $(LM_{het}L)_{IJ}$,
which determines the four-dimensional gauge couplings, is given by 
\beq
{\cal N}_{IJ}={\rm vol}(\mathbb{P}^1_b)_{het} (LM_{het}L)_{IJ} \ ,  \label{nij}
\eeq
where the indices run over the 4-dimensional gauge fields
(\ref{sixdhetaction}).

On the other hand one has in the dual type IIA on $K3$ 
\beqa
S_{IIA} &\sim& \int d^6x \sqrt{-g_{IIA}}e^{-\phi_{IIA}} 
\Bigl( R_{IIA} + (\pa_{\mu}\phi_{IIA})^2
-\frac{1}{2\cdot 3!}H_{IIA}^2\\
 && \hspace{.5cm} +\, \frac{1}{8}{\rm Tr}(\pa_{\mu}M_{IIA}L\pa_{\mu}M_{IIA}L)
-F^I_{IIA}(LM_{IIA}L)_{IJ}F^J_{IIA}
-\frac{1}{4} B\wedge { F}_{IIA}\wedge L{F}^T_{IIA}\Bigr)\nonumber 
\label{sixdiiaction}
\eeqa
with $M_{IIA}$ the symmetric moduli matrix with 
$M_{IIA}LM_{IIA}^T=L^{-1}$ for the
lattice 
intersection form
\beqa
L={\left(\begin{array}{cc}-H&0\\0&d_{ij}\end{array}\right)} .
\eeqa
Here $d_{ij}=\int_{K3}\omega_2^i\wedge \omega_2^j$ 
($i,j=1,\dots ,22$) denotes the
intersection matrix on the middle cohomology (the $\omega_2^i$ an
integral basis of harmonic two-forms).

Naively the field strength vector is given as ($C_i$ is a basis of two cycles)
\beqa
F_{IIA}^I=
{\left(\begin{array}{c}F_2\\
J_2\\
K_2^i\end{array}\right)}=
{\left(\begin{array}{c}H^{(2)}_R\\
\int_{K3}H^{(6)}_R\\
\int_{C_i}H^{(4)}_R\end{array}\right)} \label{naivefluxmapping}
 \, .
\eeqa
Here $F_2=H^{(2)}_R=dA_1$ is the Abelian field strength of the Ramond-Ramond
1-form $A_1$, $H^{(6)}_R$ is the dual Abelian field strength of 
the Ramond-Ramond 3-form $C^{(3)}$ with
$H^{(4)}_R=dC^3$ ($C^{(3)}$ is dual to a vector in
six dimensions) and the 22 field strengths $K_2^i:=dC^i_1$ correspond
to $dC^{(3)}$ where two  internal indices are on $K3$.
However, to obtain closed gauge invariant field strengths one has to 
implement the following axionic shifts \cite{Duff}
\beqa
\hat{J_2}&=&\tilde{J_2}
-J_2^ia^jd_{ij}+\frac{1}{2}F_2a^ia^jd_{ij}\nonumber\\
\hat{K_2^i}&=&J_2^i-F_2a^i=K_2^i-d(A_1a^i). 
\eeqa 
We denote by $a^i$, $C_1^i$ and $J_2^i=K_2^i+A_1da^i$ 
the parts associated with $\omega_2^i$ in the mode
decomposition for $B_2$, $C_3$ and $J_4=H^{(4)}_R+A_1\wedge H^{(3)}_{NS}$
respectively. Regarding $F_2, \tilde J_2,J^i_2$ as candidate abelian
field strengths note that neither $J_2^i$ nor
$\tilde{J_2}:=\int_{K3}*J_4$ are closed: instead one has 
$dJ_2^i=F_2da^i$ and $d\tilde{J_2}=J_2^ia^jd_{ij}$ (the latter from
the field equation $d*J_4=H_3\wedge J_4$ for $J_4$ coming from the
CS-term).

Hence the IIA field strength vector is given by the following object:
\beqa
F_{IIA}^I
={\left(\begin{array}{c}F_2\\\hat{J_2}\\\hat{K_2^i}\end{array}\right)}
={\left(\begin{array}{c}H^{(2)}_R\\
\int_{K3} H^{(6)}_R+\int_{K3}* (A_1\wedge H_{NS}^{(3)})-J_2^ia^jd_{ij}+\frac{1}{2}F_2a^ia^jd_{ij}
\\ \int_{C_i}H^{(4)}_R-d(A_1a^i)\end{array}\right)} . 
\eeqa
Note that with regard to the $\mathbb{P}^1_b$ fluxes we can use the 
naive field strength vector (\ref{naivefluxmapping}). 
Now we can identify the  type IIA $U(1)$ gauge fields with their dual
heterotic counterparts. We will first focus on the 
$\gamma$, $S$, $T$, $U$ part of the gauge group with corresponding
intersection form $H \oplus H$.
The relevant part of the $K3_{IIA}$
integral cohomology lattice consists in
one hyperbolic plane $H$ from
$H^{1,1}(K3_{IIA})=H\oplus H\oplus E_8 \oplus E_8$
related to the section (base) $\sigma$ (which is $\mathbb{P}^1_f$)
and fibre $f$ of the $K3_{IIA}$ which we assume to be elliptically
fibered
and another hyperbolic plane from $H^0(K3_{IIA})\oplus H^4(K3_{IIA})$.
For the elliptic Calabi-Yau with Hirzebruch base $F_0$ 
the intersection matrix of $\sigma$ and $f$ is indeed in the hyperbolic form. 
(For the Calabi-Yau hypersurface of degree 24 in ${\bf P}_{1,1,2,8,12}$,
which is an elliptic fibration over $F_2$, the corresponding intersection
matrix $H$ is only reached for the basis consisting of $\sigma+f$ and $f$). 
This structure $H\oplus H$ of signature $(2,2)$
is related to the coset $\frac{O(2,2)}{O(2)\x O(2)}$ and is realized
heterotically in the sector $U(1)^2_L\x U(1)^2_R$ which is associated to
the $\mathbb{T}^2_c$ sector of the heterotic compactification space.

Now utilizing the six-dimensional string/string duality one has the
following
associations of the $\gamma,S,T,U$ sector (for the Calabi-Yau with base $F_0$)
\beqa
F^\gamma_{het}  \leftrightarrow F_2, && 
F^S_{het} \leftrightarrow \hat J_2 , \nonumber\\
F^T_{het} \leftrightarrow \hat K_2^f , && 
F^{U}_{het} \leftrightarrow \hat K_2^\sigma \, .
\label{naiveU1assocminisector}
\eeqa
Note that this identification already holds in six dimensions 
and of course persists upon further compactification on 
${\mathbb P}^1_b$.

In four dimensions, one gets
analogous assertions for the electro-magnetic dual gauge field strengths
$G_\gamma^{het}$, $G_S^{het}$, $G_T^{het}$, $G_U^{het}$:
they correspond to the (Hodge) dual forms on the
(Poincare) dual cycles, i.e. in terms of the naive field strengths 
\beqa
G_{\gamma}^{het} \leftrightarrow  \int_MH^{(8)}_R , && 
G_S^{het} \leftrightarrow \int_{{\mathbb P}^1_b}H^{(4)}_R , \\
G_T^{het} \leftrightarrow \int_{{\mathbb P}^1_b\times\sigma}H^{(6)}_R , && 
G_U^{het} \leftrightarrow \int_{{\mathbb P}^1_b\times f}H^{(6)}_R ,
\label{dualgaugefields}
\eeqa
In ${\cal N}=2$ supergravity,
these electric/magnetic duality transformations,
$A\leftrightarrow\tilde A$, are given in terms of the symplectic 
transformations $X^I\leftrightarrow{\cal F}_I$, as discussed before.

\section{Fluxes on $\mathbb{P}^1$}
\label{p1fluxmap}

After the discussion of the well established type II/heterotic string duality 
without fluxes let us now investigate whether this duality survives 
turning on fluxes on the type II and on the heterotic side. We start 
considering fluxes on the 2-sphere $\mathbb{P}^1_b$.
Recall that we have identified the relevant six-dimensional gauge fields. 
Let us now map the fluxes. For that purpose we compactify further down to four dimensions
on $\mathbb{P}^1_b$, the base over which the six-dimensional space is
fibred. 
This gives  $K3_{het}\x \mathbb{T}^2_c$ on the heterotic side
or the Calabi-Yau space $M$ on
the type IIA side, respectively.
The internal 
fluxes $e_I$ will then be simply given as the internal
F-fields integrated over the  base ${\rm \mathbb{P}^1_b}$:
\beqa
e_I\, =\,\int_{
\mathbb{P}^1_b}F_I
\, .
\label{eflux}
\eeqa
Here the $F_I$ correspond to six-dimensional field strengths, where
the index $I$ ranges over those fields which survive the compactification
to four dimensions.
In addition to the $e_I$ we also like to introduce the `magnetic'
fluxes $m^I$. Together with the $e_I$ they build, 
in analogy to the magnetic/electric field
strength tensors $(G_I,F^I)$, a symplectic vector $(e_I,m^I)$.
On the type IIA side the $m^I$ will correspond to certain
forms integrated over certain cycles inside the Calabi-Yau, namely those which
do not contain $\mathbb{P}^1_b$, i.e. those inside $K3_{IIA}$, which do not have
an interpretation as fluxes of any six-dimensional gauge fields. Accordingly, 
we will see that also on the heterotic side the $m^I$ are not just 
fluxes of such six-dimensional gauge field strengths. 

As stated in the introduction, we will be interested in two types of 
terms in the Lagrangians: 
those giving the scalar potential $V$ of the superpotential $W$ induced
by the
fluxes (these are given by the kinetic terms shown below), and those
giving the relation to induced Green-Schwarz couplings;
these are essentially given by Chern-Simons terms. 

\subsection{Green-Schwarz terms, flux mapping and chiral anomalies}
\label{gsterms}
We will now compare the four-dimensional
 couplings of the (reduced) six-dimensional gauge
fields with the derivative of the type IIA (model-independent) axion
$a_{IIA}$ respectively the heterotic axion modulus $a_{ \mathbb{P}^1_b}$
related to $\mathbb{P}^1_b$.
Note that just as one has in string/string duality the well known
association
between $1/g^2_{het}$ and vol$\left(\mathbb{P}^1_b\right)_{IIA}$, 
one has equally the 
association of $1/g^2_{IIA}$ with vol$\left( \mathbb{P}^1_b\right)_{het}$ 
(see eqs.(\ref{hetiimap}) and (\ref{iihetmap})) which leads
to the corresponding association of the universal IIA B-field $B_{IIA}$ and 
the internal heterotic axion $a_{{\mathbb P}^1_b}$: 
\beqa
a_{IIA}\leftrightarrow a_{\mathbb{P}^1_b}.
\eeqa
The coupling constants are then proportional to H-fluxes of type IIA RR field
strength or to F-fluxes 
of heterotic $U(1)$ gauge fields on the common base 
${\mathbb P}^1_b$. By identifying the dual gauge 
fields on both sides we can thus map the fluxes.  

Consider the CS-improved six-dimensional kinetic
terms of the heterotic NS 2-form $B$. 
One has with $H_{het}=dB-\omega^{YM}_{CS}- \dots $ that $H^2_{het}$ 
contains the term 
$dB \cdot (A^I \wedge dA^J L_{IJ})$ for any six-dimensional gauge
field $A^I$. With the axion field in
$B=a_{ \mathbb{P}^1_b}\cdot J_{ \mathbb{P}^1_b}+\ldots$ 
this leads to the four-dimensional heterotic Green-Schwarz couplings
($e_I=\int_{\mathbb{P}^1_b} \dd A_I$)
\beqa
{\cal L}_{GS}^{het} \,   = \,
e_\gamma~  
B_{ \mathbb{P}^1_b} \wedge F^{\gamma}_{het}
+e_S~
B_{ \mathbb{P}^1_b} \wedge F^S_{het}
 +  e_T~
B_{ \mathbb{P}^1_b} \wedge F^U_{het}
+e_U~ 
B_{\mathbb{P}^1_b}\wedge F^T_{het}
\, \label{hetgs}
\eeqa
as ($T,U$) are non-diagonal w.r.t. $L_{IJ}$.

On the other hand one has on the type IIA side
the actual kinetic term $(H^{(4)}_R-2dB \wedge A^{(1)})^2$ which contains
the term $H^{(6)}_R\wedge B \wedge H^{(2)}_R$ furthermore one has the
topological term $H^{(4)}_R\wedge H^{(4)}_R\wedge B$. 
The first term leads to
\beqa
B_{IIA}\wedge \hat J_2 
\int_{\mathbb{P}^1_b}H^{(2)}_R~ 
+
B_{IIA}\wedge F_2\int_M H^{(6)}_R~\, , \label{iiagsa}
\eeqa
whereas the second term provides
\beqa
B_{IIA}\wedge \hat K_2^{j}
d_{ij} \int_{{\mathbb P}^1_b\times C_i}H^{(4)}_R~\, .\label{iiagsb}
\eeqa
For the $S$-$T$-$U$ model this leads
to
the following four-dimensional type IIA
Green-Schwarz couplings with precoefficients involving 2-, 4-, 6-flux 
\beqa
{\cal L}_{GS}^{IIA} &=&
B_{IIA}\wedge \hat J_2
\int_{\mathbb{P}^1_b}H^{(2)}_R~ 
+
B_{IIA} \wedge \hat K_2^{\sigma}
\int_{{\mathbb P}^1_b\times f}H^{(4)}_R ~ 
\nonumber\\
&& \hspace{1cm} 
+\, B_{IIA}\wedge \hat K_2^{f} 
\int_{{\mathbb P}^1_b\times\sigma}H^{(4)}_R~+
B_{IIA}\wedge F_2 \int_M H^{(6)}_R~\, . \label{iiags}
\eeqa

Now we compare the heterotic, eq.(\ref{hetgs}), and the type IIA,
eq.(\ref{iiags}), Green-Schwarz coupling, and from matching these
two effective actions we obtain the following mapping of the fluxes
(for the IIA model with Hirzebruch base $F_0$):
\beqa
e_\gamma  \leftrightarrow  \int_{M}
H^{(6)}_R,&& e_S\leftrightarrow 
\int_{{\mathbb P}^1_b}H^{(2)}_R,\nonumber\\
e_T \leftrightarrow  \int_{\mathbb{P}^1_b\x f}
H^{(4)}_R, && 
e_U\leftrightarrow \int_{\mathbb{P}^1_b\x
\sigma}H^{(4)}_R\, ,
\label{emap}
\eeqa
of course in consistency with the $\mathbb{P}^1_b$-integrated version of 
\ref{naiveU1assocminisector}.

Note that the flux $e_S$ corresponds on the type IIA
side to the vacuum expectation value of the 2-form $H^{(2)}$ over
${\mathbb P}^1_b$. This is in contrast to the fluxes $e_T$ and $e_U$
(and for the further gauge fields - see section (2.4)) which are determined
by the vevs of $H^{(4)}$. However this observation is just a reflection of the
fact that the weakly coupled heterotic $S$-gauge field
corresponds to the period ${\cal F}_1$, as explained before.

Now let us come to the dual magnetic fluxes
$m^I$. They correspond in the
type IIA theory to the (Hodge) dual forms on the 
(Poincare) dual cycles. Hence we obtain the following identifications
\beqa
& &m^{U}  \leftrightarrow 
\int_{f}H^{(2)}_R,\quad
m^S \leftrightarrow 
\int_{K3_{IIA}} H^{(4)}_R,\quad\nonumber\\
& &m^T \leftrightarrow 
\int_{\sigma}H^{(2)}_R,\quad
m^{\gamma} 
\leftrightarrow \int_{pt.}H^{(0)}_R\, .\label{mfluxes} 
\eeqa
Starting from the ten-dimensional type IIA action with terms 
$H^{(6)}_R\wedge B\wedge H_R^{(2)}$ and $H^{(4)}_R\wedge H^{(4)}_R\wedge B$,
the dual fluxes generate the following `dual' Green-Schwarz terms in four
dimensions:
\begin{equation}
{\cal L}_{GS}^{dual}=m^S~B_{IIA} \wedge G_S
+m^U~B_{IIA}\wedge G_T+m^T~B_{IIA}\wedge G_U\, . 
\end{equation}
Note that the still missing  GS-coupling 
\begin{equation}
{\cal L}_{GS}^{dual}=m^\gamma ~B_{IIA}\wedge G_\gamma\, ,
\label{mgs}
\end{equation}
cannot be derived from the two ten-dimensional terms written above,
but requires in ten dimensions a coupling of the form
$H^{(8)}\wedge B$. This term is not present in the effective action
of the `normal' type IIA superstring, but it precisely arises in the massive
IIA theory which was discussed in \cite{Romans:1986tz}.
Specifically, in the massive IIA theory there is a cosmological constant term,
$\Lambda\sim (m^\gamma)^2$, and the 2-form field strength has to be
modified in the following way:
\beqa
H^{(2)}_R=dA_1+2m^\gamma B_{IIA}\, .
\eeqa
Then the term in  eq.(\ref{mgs}) arises in the kinetic term $(H^{(2)}_R)^2$.

Note that there is no integration on the base $\mathbb{P}^1_b$ in any of these
integrals (\ref{mfluxes}) involving the fluxes $m^I$.
Therefore the fluxes $m^I$ already exist in six dimensions
as integral, topological numbers on the type IIA side.
In fact, decompactifying ${\mathbb P}^1_b$ in type IIA, the fluxes
$m^I$ are already present in six dimensional IIA compactification on
$K3$. All these fluxes arise in massive type IIA supergravity
on $K3$ 
due to duality transformations
from $m^\gamma$, as it was recently discussed in \cite{Haack:2001iz}.

However on the heterotic side the interpretation of the magnetic fluxes
$m^I$ is not obvious, since the corresponding
integrals cannot be defined (but see \ref{dualinterpr} below) Therefore the
heterotic/type IIA string-string duality breaks down in the presence
of the magnetic fluxes $m^I$.
In fact, 
the problematical point to match these two occurrences of Poincare-duality
is the following: the  electro-magnetic dual
field strengths
$G_\gamma$, $G_S$, $G_T$, $G_U$
are defined entirely in {\em four} dimensions; this is
in manifest contrast to the original gauge fields which are survivors
from a dimensional reduction. If one wants to contemplate on
${\mathbb P}^1_b$-fluxes of the electro-magnetic duals they should
exist already in {\em six} dimensions. 
In conclusion there are no fluxes of the electro-magnetic duals of
the four-dimensional gauge fields. In other words,  
the four IIA fluxes listed in eq.(\ref{mfluxes})
with no support on the ${\mathbb P}^1_b$ cannot be interpreted as heterotic
F-fluxes of six-dimensional gauge fields on ${\mathbb P}^1_b$.

Let us briefly discuss the action of the duality symmetries, being certain
symplectic transformations, on the fluxes $e_I$ and $m^I$. First,
the perturbative, heterotic T-duality transformations act within
the fluxes $e_I$ and $m^I$ but do not mix them into each other.
For example, the transformation $T\rightarrow 1/T$ has the following action on
the fluxes: $e_S\leftrightarrow e_U$, $e_\gamma\leftrightarrow e_T$,
$m^S\leftrightarrow m^U$, $m^\gamma\leftrightarrow m^T$ as inspection of the 
scalar potential (\ref{stuscalarpot}) shows.
On the other hand, non-perturbative S-duality transformation exchange
some of the fluxes $e_I$ with some $m^I$. Therefore, in the heterotic
string, the absence of the fluxes $m^I$ breaks these S-duality
transformations. In the type IIA string, non-perturbative
duality transformations which
exchange $e_I$ with $m^I$ are in principle possible and depend on
the details of the geometry.
For example, as discussed in 
\cite{Klemm:1995tj,Cardoso:1996zu},
the $S-T-U$ models discussed here possess a non-perturbative exchange
symmetry
$S\leftrightarrow T$. In the model with Hirzebruch base $F_0$, this symmetry
has a very simple geometric interpretation, namely it just reflects the
freedom of exchanging the two ${\mathbb P}^1$'s of $F_0$, i.e.
${\mathbb P}^1_b\leftrightarrow\sigma$.
Using the fluxes in eqs.(\ref{emap}) and (\ref{mfluxes})
the $S\leftrightarrow T$ symmetry has the following natural
action on the fluxes: 
\beqa
e_T\leftrightarrow m^S, \quad e_S\leftrightarrow m^T \ , \label{dualinterpr}
\eeqa
and all other fluxes are unchanged.
One gets of course similar mappings if one assumes a full triality
among the moduli $S$, $T$ and $U$ (see e.g. 
\cite{LopesCardoso:1996qa,Gregori:2000sn}).

Now, at the end of this chapter let us discuss the question of anomalies
in relation to the ${\mathbb P}^1_b$-fluxes.
The net number of chiral fermions, the index of the Dirac operator, 
which is proportional to the triangle $U(1)^3$-anomaly is given by  
\beqa
n_+-n_- = \int_{\mathbb{T}_c^2\times K3_{het}}{ {\rm ch}\left(F\right) \wedge 
\hat A (R) }  ,
\eeqa
where $\hat A(R)$ and ch$(F)$ are the $\hat A$ genus of the tangent bundle and
the Chern character of the gauge bundle. The index 
is zero due to the simple fact that on the `constant' $\mathbb{T}_c^2$
$F$ and $R$ both vanish. This reflects the statement that we can figure the
heterotic compactification as a trivial dimensional reduction on a torus after
compactifying to six dimensions on K3$_{het}$, which does not produce any
chiral fermions in four dimensions. 
This is still completely consistent with the appearance of Green-Schwarz
couplings. A contribution to the triangle anomaly would require the coupling 
$a_{{\mathbb P}^1_b} F\wedge F$. In general, this term can be obtained from
the ten-dimensional GS-term $B\wedge F^4$ via dimensional reduction.
However for the case of ${\mathbb P}^1_b$-fluxes we obtain
\begin{equation}
{\cal L}=a_{{\mathbb P}^1_b}\, F^I\wedge F^I~\int_{{\mathbb T}^2_\sigma
\times {\mathbb T}^2_f}F\wedge F\, ,
\end{equation}
which vanishes since $F$ has only support on ${\mathbb P}^1_b$. 

The gauge symmetry not being anomalous, the GS-term $B_{\mathbb{P}_B^1} \wedge
F$ nevertheless generates a longitudinal
component for the $U(1)$ gauge fields. Since there is only
one B-field, which couples to the gauge fields via the GS-term,
namely the internal B-field  $B_{{\mathbb P}^1_b}$ in the heterotic string
or the universal B-field $B_{IIA}$ in the type IIA string,
only one linear combination of $U(1)$ gauge fields will become massive,
where the exact linear combinations depends on the fluxes turned on.
In the heterotic string this mass term comes   
from the ten-dimensional kinetic term of the 
CS-improved NSNS 3-form field strength, which exists at string tree-level. 
The generated mass for the $A^I_\mu$ is
proportional to $e_I^2/{\rm vol}({\mathbb P}^1_b)^2$. 
However using the 
heterotic/type IIA duality relation (\ref{iihetmap}),
it follows that these gauge boson mass terms arise only
at the one-loop level in the type IIA compactifications.

\subsection{The superpotential couplings}
\label{hetsupo}

Turning on non-trivial fluxes induces a potential
for the vector multiplet scalar fields $z^A=X^A/X^0$.
Quite generically, this scalar potential is lifting the vacuum
degeneracy
in the vector multiplet moduli space and is also breaking space-time
supersymmetry. In the heterotic case the scalar potential originates
from the gauge kinetic term 
eq.(\ref{sixdhetaction}) in six dimensions, which is determined
by the scalar field matrix $(LM_{het}L)_{IJ}$.
Specifically, the four-dimensional scalar potential $V$ 
is obtained by replacing
in eq.(\ref{sixdhetaction}) the gauge field strength tensors $F^{het}$
by their corresponding internal electric fluxes $e_I$.
Using eqs.(\ref{magneticg}), (\ref{nij}) and (\ref{eflux}) this 
leads to following expression written in terms of four-dimensional variables:
\beqa
V={1\over {\rm vol}({\mathbb{P}^1_b})_{het}}~e_I({\rm Im}\hat{\cal N})^{IJ}e_J\, ,
\label{scalarpot}
\eeqa
where the coupling constants $\hat{\cal N}_{IJ}$ are computed using
the period vector $(\hat X^I,\hat{\cal F}_J)$ which was obtained from
the exchange $\hat X^1={\cal F}_1$, $\hat{\cal F}_1=-X^1$.
Using the well-known supergravity
relation between $W$ and $V$, this scalar
potential corresponds  to 
the following superpotential
\beqa
W=e_\gamma X^0-e_S\frac{\partial {\cal F}}{\partial X^1}+ e_iX^i\, ,
\eeqa
where $S=-iX^1/X^0$ and the fields $\phi^i=-iX^i/X^0$ are the remaining 
moduli. In heterotic perturbation theory, the prepotential ${\cal F}$
depends only linearly on $S$, and hence $W$
does not depend on $S$; therefore the $S$-field is not fixed by the
minimization condition $W=0$ and $d_iW=0$ in perturbation
theory. Only non-perturbatively, after
instanton corrections in $e^{-S}$, ${\cal F}$ and hence also $W$ will
depend on $S$, which will then be fixed by the minimization conditions.
For the case with only four vector fields $A_\gamma$, $A_S$, $A_T$ and
$A_U$
with corresponding prepotential eq.(\ref{hetprepot})
the couplings $\hat{\cal N}^{IJ}$ are easily computed
using
eq.(\ref{goods}), and we get
the following heterotic scalar
potential (for real values of the moduli):
\beqa
V={1\over {\rm vol}({ \mathbb{P}^1_b})_{het} }{(e_\gamma)^2+(e_S)^2 T^2U^2+(e_T)^2 T^2+(e_U)^2U^2
\over 2STU}\, ,\label{stuscalarpot}
\eeqa
which corresponds to the following superpotential cf. \cite{CKLTh}:
\beqa
W=e_\gamma+e_S TU+ie_TT+ie_UU\, .\label{STUsupo}
\eeqa

On the type IIA side the flux induced scalar potential
originates
from the  kinetic terms $(H^{(n)}_R)^2$ in ten dimensions. 
Alternatively we can
also use the derivation \cite{VT}
of the scalar potential in the type IIB
mirror compactification. There the scalar potential comes from the 
the term $H^{(3)}_R \wedge * H^{(3)}_R$ in the ten-dimensional
effective action. After dimensional reduction the following expression
is obtained \cite{VT}:
\beqa
V=-(2{\rm Im}\tau)^{-1}\biggl(m^I({\rm Im}{\cal N})_{IJ}m^J+
(e_I+m^J\bar{\cal N}_{IJ})({\rm Im}{\cal N})^{IK}(e_K+m^J{\cal N}_{KJ})
\biggr)\, .
\eeqa
The $e_I$-part of eq.(\ref{scalarpot}).
In type IIA, $V$ can be derived from the following superpotential
\beqa
W = \sum_{n=0}^3\int_W H_R^{(2n)}\wedge J^{3-n}= 
e_I~ X^I - m^I~
{\cal F}_I\, .
\eeqa 
With the six-dimensional mode decompositions 
$H^{(4)}_R=H_2^i \wedge \omega_i $ (apart from contributions with a 0-
respectively 4-form on $K3_{IIA}$) and $J=X^i \cdot \omega_i$
(where $X^i=a^i+i{\rm vol}(C_2^i)_{IIA}$)
on $K3_{IIA}$ one then obtains that 
\beqa
\int_M H^{(4)}_R\wedge J= X^j d_{ij} \int_{\mathbb{P}^1_b}H_2^i =
X^j d_{ij} \int_{\mathbb{P}^1_b\x C_2^i}H^{(4)}_R =e_iX^j d_{ij}
\eeqa
and furthermore 
\beqa
e_0=\int_MH^{(6)}_R=e_\gamma \, .
\eeqa
$H^{(2)}_R\wedge \frac{1}{2}J^2$ contains the contribution
\beqa
m^1 \int_{K3_{IIA}}\frac{1}{2}J^2=
\int_{\mathbb{P}^1_b}H^{(2)}_R  \int_{K3_{IIA}}\frac{1}{2}J^2=
e_S\frac{\pa {\cal F}}{\pa X^1}\, .
\eeqa
So in total one gets  the following
superpotential
\beqa
W&=&e_iX^j d_{ij}+ e_0X^0 -
m^1\frac{\pa {\cal F}}{\pa
X^1}\nonumber\\
 &=&ie_TT+ie_UU+ e_\gamma + 
e_S TU\, ,
\eeqa
where we have used the prepotential ${\cal F}$ in
eq.(\ref{hetprepot}) and $X^0=1$ in the second line of this equation. 
The agreement of this equation with heterotic superpotential
eq.(\ref{STUsupo})
shows how the superpotential can be directly transported from
type IIA to the heterotic side, whereas above we compared these terms
by explicitely evaluating the associated scalar potentials on both
sides independently.

\subsection{Further gauge fields}

It is straightforward to extend these associations to the case
that one has heterotically not only the $T$ and $U$ fields from the
$g$ and $B$ sector but also further gauge fields, called $V$. 
So if one of the 16 $U(1)$'s
survives the fibration down to 4 dimensions 
one can consider its associated flux. Such a model will be 
dual to type IIA on a 
Calabi-Yau three-fold with a larger $h^{(1,1)}$ reflecting the
enhanced number of vector multiplets \cite{MV,CF}. Its 
$K3_{IIA}$ fibre has a correspondingly higher Picard number which
indicates the existence of at least one further (beyond the $\sigma$
and $f$) algebraic 2-cycle $C_V^2$ whose
cohomology class in the $K3_{IIA}$ is an integral (1,1) class and which
exists generically in all $K3_{IIA}$ fibers. This gives a new 2-cycle
for the Calabi-Yau and a new 4-cycle ${\mathbb P}^1_b\;\x C^2_V$ 
from the adiabatic extension over the base ${\mathbb P}^1_b$.   
Of course, the Wilson lines then
correspond to the Cartan sub lattice of the $E_8\oplus E_8$ part of the
$K3_{IIA}$ middle cohomology.

The heterotic gauge field (existing already in six dimensions)
with field strength $F_{het}^V$  
maps to the the RR 3-form $C^{(3)}$ reduced 
on the new 2-cycle $C_V^2$, i.e.
\beqa
F_{het}^V\leftrightarrow F^V_{IIA}=\int_{C_V^2}H^{(4)}_R\, ,
\eeqa
whereas the electro-magnetic dual field strengths $G^{het}_V$ 
(existing only in four dimensions)
corresponds again to
the dual 5-form $C^{(5)}$ on the Poincare dual 4-cycle $C_V^4$, 
i.e.
\beqa
G^{het}_V\leftrightarrow G_V^{IIA}=
\int_{C_V^4}H^{(6)}_R\, .
\eeqa
Given these identifications of the heterotic versus type IIA $U(1)$
field strengths one has immediately the following 
associations of fluxes on $\mathbb{P}_b^1$ 
\beqa
e_V=\int_{{\mathbb P}^1_b}F^V_{het}&\leftrightarrow &
\int_{{\mathbb P}^1_b\x C_V^2}H^{(4)}_R\, , \nonumber\\
m^V&\leftrightarrow & \int_{({\mathbb P}^1_b\x C_V^2)^{\bot}}H^{(2)}_R\, .
\eeqa

\section{$\mathbb{T}^2$-Fluxes}
\label{hett2fluxes}

We now turn to heterotic fluxes on the constant torus
$\mathbb{T}^2_c$. Therefore, we figure the
compactification space $K3\x \mathbb{T}^2_c$
to be build up in the other order B: One decompactifies first to six
dimensions on $K3$ and then further to four dimensions on
$\mathbb{T}^2_c$. Any abelian 
gauge field $A$ that survives the compactification on K3 may
also be endowed with a flux $f$ on ${\mathbb T}^2_c$, 
\beqa
f =  \int_{\mathbb{T}^2_c}{F}
\in
\mathbb{Z}\, .\label{fluxf0}
\eeqa 
There is no straightforward way to interpret these fluxes on the dual type IIA
side, but we would like give some indications. 
 
First recall that for the heterotic ${\mathbb P}^1_b$ fluxes in 
eq.(\ref{eflux}) the vacuum expectation value of the internal 
magnetic field is 
inversely proportional to the ${\mathbb P}^1_b$-volume,
\begin{equation}
F_{{\mathbb P}^1_b}\sim \frac{1}{{\rm vol}({\mathbb P}^1_b)_{het}}\, .
\label{volp}
\end{equation}
On the type IIA side ${\rm vol}({\mathbb P}^1_b)_{het}^{-1}$ corresponds 
to the
square of the type II coupling constant (see eq.(\ref{iihetmap})), such that
\begin{equation}
F_{{\mathbb P}^1_b}\sim  g^2_{IIA}\, .
\end{equation}
Using the same arguments the heterotic ${\mathbb T}^2_c$ fluxes scale like
\begin{equation}
F_{{\mathbb T}^2_c}\sim\frac{1}{{\rm vol}({\mathbb T}^2_c)_{het}}\, ,
\end{equation}
and keeping the overall volume of $K3\times {\mathbb T}^2_c$ as well as
the volume of ${\mathbb T}^2_f$ fixed we arrive at
\begin{equation}
F_{{\mathbb T}^2_c}\sim {\rm vol}({\mathbb P}^1_b)_{het} \sim 1/g^2_{IIA}\, .
\end{equation}
So in the dual type II description these fluxes are large for weak coupling,
i.e. they are of non-perturbative nature. A similar feature is
encountered in type IIB Calabi-Yau compactifications in the presence of
NS-fluxes, where  the flux vectors are 
$e=e_R+\tau e_{NS}$, $m=m_R+\tau m_{NS}$ , 
which originate from vacuum expactation
values of $H^{(3)}=\tau H^{(3)}_{NS}+H^{(3)}_R$ \cite{VT}\footnote{In \cite{Vafa:2000wi} 
it was argued that the $H^{(3)}_{NS}$ fluxes correspond on the type IIA side
to fluxes of an NS 4-form field strength.}.

As opposed to the earlier case of the $\mathbb{P}_b^1$ fluxes we now expect to
obtain a spectrum of chiral fermions in four dimensions together with an
anomalous contribution to the chiral gauge anomaly. 
The index theorem for the Dirac operator relates the
net number of chiral fermions to the magnetic flux on the torus
\beqa \label{inddir}
n_+ - n_- = 
\int_{K3}{\hat A(R)} \int_{\mathbb{T}_c^2}{F} 
\sim f\int_{K3}{{\rm tr}\, R^2}
. 
\eeqa
But now we also face the presence of appropriate Green-Schwarz couplings to
cancel the anomaly. The kinetic term of the NSNS 3-form $H_{het}$ decomposes
into  
\beqa
H_{het} \wedge *H_{het} = 2 dB \wedge * \left( AdA \right) +
A dA \wedge * \left( AdA \right) +\ \cdots
\eeqa
The second term is the term that gives a positive mass 
to the particular gauge boson via $f^2\, A_\mu A^\mu$. 
Note that this mass square is proportional to the square of the flux. 
The first term is a contribution to
the Green-Schwarz coupling $B\wedge F$, which simply is of the following form
in four dimensions:
\begin{equation}
{\cal L}_{GS}= f ~B_{{\mathbb T}^2_c}\wedge F\, .
\end{equation}
Via the index theorem, the coupling constant is proportional to the net number
of chiral fermions, thus appropriate to cancel the anomaly induced 
by the triangle diagrams with fermion loops. To complete the Green-Schwarz 
couplings needed for the anomalous tree diagram the ten-dimensional 
term $B\wedge F^4$  leads in
four dimensions to a coupling
\begin{equation}
{\cal L}=a_{{\mathbb T}^2_c}~F\wedge F\int_{K3}F\wedge F\, ,
\end{equation}
which is non-vanishing and independent of the flux. Using the Bianchi identity
\beqa
\int dH_{het} = \int \left( {\rm tr}\, F^2 - {\rm tr}\, R^2 \right) = 0 
\eeqa
the total GS coupling is proportional to the index and the 
two terms precisely surfice to cancel the non-vanishing $U(1)$ anomaly. 

\section{Tachyons due to fluxes}

It is well known that it can lead to unstable vacua when
Yang-Mills gauge theories are compactified to lower dimensions with 
magnetic fluxes on the internal space \cite{NO78}. 
The internal components of charged gauge bosons become scalars
whose masses are modified due to their coupling to the internal magnetic
field and thus may become negative eventually. Then the vacuum shifts to a new
groundstate with a condensate of these tachyons. The analysis can be
performed on  purely field theoretical grounds by considering six-dimensional
Yang-Mills theory compactified to four dimensions. It equally applies to both,
the type IIA and heterotic string models. 

First let us dicuss tachyons due to $\mathbb{P}^1$-fluxes. 
Later we will consider this mechanism in the more familiar case of a toroidal
compactification of gauge fields with magnetic flux. 
There we will also explicitly show that the tachyon potential that
derives from the gauge kinetic term tr$F^2$ is stabilized by a quartic term,
the entire potential taking the form of a D-term.  
But, at first sight surprising, tachyons can also appear for a
compactification on a $\mathbb{P}_b^1$. One might expect that the absence of
harmonic 1-forms, which implies the absence of massless scalars in the
dimensional reduction of a gauge field on $\mathbb{P}_b^1$, also excludes
their appearance when magnetic flux is present on the sphere. 
However, the analysis of the spectrum of the Laplacian on a sphere acting on
internal components of charged vector fields shows that it 
indeed does have negative eigenvalues as well \cite{Dvali,Daemi}. 
Formally, one needs to consider sections in the tangent bundle twisted by the
flux, which may be trivial even if the tangent bundle itself is not.  
The flux being a discrete parameter the existence of tachyonic modes 
is not in contradiction with  being massive Kaluza-Klein excitations in
its absence. An important feature is that the modification of 
their masses is essentially given by $\delta M^2 \sim sqF$, $s$ denoting the
internal spin, $q$ the charge and $F$ the flux, a 
linear dependence on the flux $F$. 
Hence, the spacing of mass levels of KK states and the shift induced by the
flux is comparable, both being proportional to $1/{\rm
  vol}(\mathbb{P}_b^1)$. While this explains how the mass shift can let the
massive KK modes jump to negative eigenvalues, it also raises the general
question if the approach based on the effective
action of fields massless in
the absence of flux is appropriate at all. The two scales being equal one would
have to include all higher KK modes from the beginning, because all of them
could come down to zero mass when coupling to appropriate internal
fluxes. Thus, the starting point of our analysis appears to be ruined.  
This problem does not occur as long as we consider only abelian gauge
symmetries without charged gauge bosons.

Another more exotic situation where charged scalar fields have negative masses 
can be met at very special points of the moduli space. When the
compactification space degenerates and certain cycles shrink to zero size,
black holes may condense and need to be included in the effective action even
at small coupling. 
Let us look at the well known example of the conifold singularity of type IIB, 
which occurs at the co-dimension one locus,
where in the Calabi-Yau space a cycle $A_1$ with the topology of
$S^3$ vanishes, while the remaining cycles stay finite.
More precisely the Calabi-Yau space $M$ exhibits a nodal singularity, i.e.
it is described locally by the eq. $\sum_{i=1}^4 \epsilon_i^2=\mu$. For
$\mu\rightarrow 0$ the real part of this local
equation describes the vanishing $S^3$. In the vicinity of a conifold point,
$X^1=\int_{A_1} \Omega\rightarrow 0$, an additional hypermultiplet, the ground 
state of a singly wrapped $3-$brane
around the $A_1$, with mass proportional to $|X^1|$ becomes light
\cite{stromon}. It is charged with respect to the
$U(1)^{N_V}$ gauge symmetry of the vector multiplets. It 
corresponds to a magnetic monopole or dyon in the effective gauge theory.

Consider the case where the flux $e_1$ which is aligned to the vanishing
cycle of the conifold is turned on. In type IIB the corresponding superpotential is
\begin{equation}
W=e_1\mu\, +\, \mu\phi\tilde\phi \, ,
\end{equation}
where we have set
\be
\mu={X^1\over X^0} \ .
\eq
The supersymmetric, stationary points of the 
corresponding scalar potential are at
$W=0$ and $dW=0$, which leads  to $\mu=0$ and in addition to the condensation
of the hyper multiplets,
$\phi\tilde\phi=-e_1$, as discussed in \cite{VT}.
In fact the supergravity scalar potential  in the vicinity of
the conifold point \cite{PolStro,CKLTh},
\begin{equation}
V(\mu)=|W_\mu|^2e^K K^{-1}_{z\mu\bar\mu}=-{e_1^2\over \log |\mu|^2},
\label{vconi}
\end{equation}
is stable in the $\mu$-direction and has 
a supersymmetry preserving minimum at $\mu=0$ with $v=0$.
On the other hand due to the non-vanishing flux $e_1$, two of
the real scalars of $\phi$,
$\tilde\phi$ will become real massive scalars whereas the remaining two
scalars will become tachyons.
Explicitly, besides eq.(\ref{vconi})
the scalar potential will also contain the term
\begin{equation}
V(\phi,\tilde\phi)=e_1(\phi\tilde\phi +{\rm c.c.})=
2e_1(\phi_1\tilde\phi_1-\phi_2\tilde\phi_2)\, ,
\end{equation}
where $\phi=\phi_1+i\phi_2$ and $\tilde\phi=\tilde\phi_1+i\tilde\phi_2$.
So the fields $\phi_2$ and $\tilde\phi_2$ are tachyons with negative
mass square $M^2=-2e_1$.

Actually, the conifold and other known examples are no good
examples for ${\mathbb P}^1$-fluxes in the context of the dual 
heterotic/type IIA  models discussed above. In IIA language the entire CY shinks
to zero. This however corresponds to a magnetic
flux, dual to the flux of $A_\gamma$, i.e. not a ${\mathbb P}^1$-flux.

For the toroidal case one can demonstrate the appearance of tachyonic
charged scalars more explicitly by the dimensional reduction of the gauge
kinetic term with background flux. We therefore assume
that a whole non-abelian $SU(2)_V$ remains unbroken in six dimensions.
Then the six-dimensional kinetic term 
\beqa
{\cal L}^V_{\rm YM} = \frac{1}{g^2_{\rm YM}}{\rm tr}F \wedge *F\, ,
\quad F^a = dA^a + f^{abc} A^b\wedge A^c 
\eeqa
decomposes according to   
\beqa \label{ymdecomp}
F^a\wedge * F^b &=& dA^a\wedge *dA^b + 2 f^{bcd} dA^a \wedge * \left( A^c
\wedge A^d \right) \non 
&& \hspace{3cm} + f^{acd} f^{bef} A^c \wedge A^d \wedge *\left( A^e\wedge A^f
\right) . 
\eeqa
With the abelian flux $f$ for $A^0$ as in eq.(\ref{fluxf0})
the second term is a mass term for the
internal components of the six dimensional charged vector fields
$A^a_{5,6}, a\not=0$. Changing to complex coordinates 
\beqa
A_\pm^a = \frac{1}{\sqrt{2}} \left( A_5^a \pm i A_6^a \right) 
\eeqa
on the torus and to the Cartan basis in group space, the mass
matrix becomes diagonal
\beqa
2 f^{0bc} A_5^b A_6^c = -\frac{1}{2} f^{0bc} \left( \vert A_+^b +
i A_+^c
  \vert^2 - \vert A_-^b + i A_-^c \vert^2 \right) 
\eeqa
such that we get two real massive scalars and two tachyons. 
The quartic term in (\ref{ymdecomp}) stabilizes the tachyons, it just 
completes the square to get a D-term potential
\beqa
{\rm tr}\left( F_{56} \right)^2 = 
\left( f - \frac{1}{2} f^{0bc} \left( \vert
    A_+^b + i A_+^c \vert^2 - \vert A_-^b + i A_-^c \vert^2 \right)
\right)^2 .  
\eeqa
This is an agreement with the potential derived in
\cite{Anton},\cite{Bachasal}, which was obtained by computing string
corrections to a quantum mechanical mass formula 
\beqa \label{tachmass}
M^2= (2n+1) \vert qf \vert  -2sqf ,
\eeqa
where $n$ is the angular momentum quantum number, the Landau level, and
$s$
the internal spin. A characteristic property of the mass square 
shift described by this formula is the linear dependence 
on the flux $f$, which appears as a generic
property of D-term induced masses. 
The mass spectrum thus obtained satisfies the condition 
\beqa
{\rm Str}M^2 =0
\eeqa
for spontaneous supersymmetry breaking. For $s=\pm 1$ the above mass 
formula produces one positive and one negative mass-square scalar field.  
Concerning the spectrum of massless fermions the mass shift affects one
chirality with $s=\pm 1/2$ to become massive, the opposite chirality to
stay
massless, such that the resulting four-dimensional spectrum will be
chiral. 
While the expression (\ref{tachmass}) has been deduced from
more heuristic arguments, it can be reproduced for a compactification of
the
heterotic string on $\mathbb{T}^6$ with 
a background $U(1)$ flux from the exact CFT
treatment of \cite{Tseytlin}. It appears as the leading term of the
exact mass
formula for small values of the flux. 

\section{Conclusion}
\label{conclusions}

In this paper we have seen that in general
F- and H-fluxes in heterotic and type II 
compactifications apparently break the string-string duality symmetry between
heterotic/type II pairs which were dual to each other before turning on
these fluxes. 
However for a subset of fluxes the string-string duality still holds,
namely for those F- and H-fluxes which have support on the two-sphere
${\mathbb P}^1$ which is common to both string compactifications.
One should ask what is the reason for the breakdown of the string-string
duality symmetry by the other fluxes, or whether one can reconcile
the string-string duality symmetry by turning on new fluxes which are
so far not yet investigated. But unfortunately we cannot find any real trace
for such a possibility in our discussion.

Another interesting question is the vacuum structure of heterotic and type II
compactifications in the presence of fluxes. As discussed in
\cite{PolStro,Mich,VT,Mayr,CKLTh} in the context of the effective
supergravity action vacua with completely
unbroken ${\cal N}=2$ supersymmetry are possible at certain degeneration
points in the moduli space (conifold points, large volume limit, etc);
otherwise supersymmetry will be completely broken, and it seems to be no
room for partial ${\cal N}=2\rightarrow{\cal N}=1$ supersymmetry breaking
in the flux induced supergravity action. However this situation changes
in the rigid field theory limit where  certain so far dynamical fields are
frozen. This field theory limit can be conveniently described by
replacing the compact type II Calabi-Yau spaces by their non-compact 
counterparts which can be then utilized to compute the flux induced 
field theory ${\cal N}=1$ superpotentials for the corresponding
Yang-Mills gauge theories without \cite{Vafa:2000wi} 
and also with matter fields 
\cite{Cachazo:2001jy}. A nice way to understand the related large N duality
between D-branes and H-fluxes, which is
based on topological transitions in the non-compact type II
Calabi-Yau spaces,
was provided in terms of geometric transitions in M-theory on
non-compact spaces with $G_2$-holonomy \cite{Atiyah:2000zz}.
In the light of this result it would be also interesting to see whether
also type II compactifications on compact Calabi-Yau spaces with H-fluxes,
or their heterotic counterparts, can be lifted to some geometric
M-theory
compactifications. In case one would succeed to have partial
supersymmetry breaking from ${\cal N}=2\rightarrow {\cal N}=1$ even
in the compact case, M-theory should be compactified on a compact
seven-dimensional space with $G_2$-holonomy.
However for the generic
case of complete supersymmetry breaking by the fluxes, the
situation looks much more complicated.

\section*{Acknowledgements}

We would like to thank Jan Louis and Stefan Theisen
for many valuable discussions and Seif Randjbar-Daemi 
for helpful communication. This work is supported by the European 
Commission RTN programe HPRN-CT-2000-00131, by GIF - the German-Israeli 
Foundation for Scientific Research and by the Studienstiftung des 
deutschen Volkes as well as the DFG.

\end{document}